\documentclass[reprint,superscriptaddress,amsmath,amssymb,aip]{revtex4-1}
\usepackage{amssymb,amsmath,amsthm,wrapfig,esint,afterpage,marginnote,afterpage,lipsum,color,tabularx}
\usepackage{bm}
\usepackage{lineno}
\usepackage{siunitx}
\usepackage{graphicx}
\usepackage{lipsum}
\usepackage{color,sidecap,fancyhdr}
\usepackage[dvipsnames]{xcolor}
\usepackage{mathpazo}
\usepackage{textcomp}
\usepackage{nicefrac}
\usepackage{xfrac}
\usepackage[sort&compress]{natbib}
\usepackage{dcolumn,bm}
\usepackage{hyperref}
\usepackage{cleveref}
\usepackage{mathtools}
\usepackage{epsfig}
\usepackage{epstopdf}
\usepackage[authormarkuptext=name]{changes}
\usepackage{changes}
\definechangesauthor[color=blue, name={AY}]{AY}
\newcommand{\colr}[1]{{\color{red} #1}}
\newcommand{\colb}[1]{{\color{blue} #1}}

\newcommand{\LT}{L_{\text{T}}}
\newcommand{\kT}{k_{\text{T}}}
\newcommand{\LH}{L_{\text{p}}}
\newcommand{\rT}{\rho_{\text{T}}}
\newcommand{\rH}{\rho_{\text{H}}}
\newcommand{\rA}{\rho_{\text{A}}}

\newcommand{\vP}{{\bf P}}

\begin{document}
	
	\title{The nonlinear initiation of side--branching by activator-inhibitor-substrate (Turing) morphogenesis}
	
	\author{Arik Yochelis}\email{yochelis@bgu.ac.il}
	\affiliation{Department of Solar Energy and Environmental Physics, Blaustein Institutes for Desert Research (BIDR), Ben-Gurion University of the Negev, Sede Boqer Campus, Midreshet Ben-Gurion 8499000, Israel}
	\affiliation{Department of Physics, Ben-Gurion University of the Negev, Beer-Sheva 8410501, Israel}

	\received{\today}
	
	\begin{abstract}
		An understanding of the underlying mechanism of side--branching is paramount in controlling and/or therapeutically treating mammalian organs, such as lungs, kidneys, and glands. Motivated by an activator-inhibitor-substrate approach that is conjectured to dominate the initiation of side--branching in pulmonary vascular pattern, I demonstrate a distinct transverse front instability in which new fingers grow out of an oscillatory breakup dynamics at the front line, without any typical length scale. These two features are attributed to unstable peak solutions in 1D that subcritically emanate from the Turing bifurcation and that exhibit repulsive interactions. The results are based on a bifurcation analysis and numerical simulations, and provide a potential strategy toward developing a framework of side--branching also of other biological systems, such as plant roots and cellular protrusions.
	\end{abstract}
	
	\maketitle
	
	{\noindent {\bf
			Reaction-diffusion type models are widely employed as a laboratory for the study of pattern formation in spatially extended systems that are driven far from equilibrium, ranging from biological and ecological realizations to technological applications. In the biomedical context, multi-variable reaction-diffusion models are often referred to as activator-inhibitor-substrate systems. While complete modeling of the morphogenesis of organs is difficult to realize, uncovering partial mechanisms responsible for it remains of utmost importance for understanding functional aspects and application design. This study demonstrates a distinct type of transverse front instability and is related to the nucleation of side--branches. As such, the results open new vistas that are likely to be relevant in a variety of biological and ecological systems.
		}
		
		\section{Introduction}
		Several essential organs in mammals~\cite{ochoa2012branching,hannezo2019multiscale}, such as the lungs, kidneys, pancreas, and mammary glands, self--organize in tree-like branched architectures, a form that enables the exploitation of a large active surface area while preserving a small volume~\cite{caduff1986scanning,roth2005neonatal,warburton2008order,metzger2008branching,lu2008patterning,affolter2009tissue,costantini2010patterning,yao2011matrix,little2012mammalian,iber2013control,davies2015epithelial,hannezo2017unifying}. Two types of basic processes are noticeable in the structural development with respect to locations along the ``mother'' branch: (\textit{i}) Tip splitting, where the ``mother'' branch deforms into ``daughter'' branches~\cite{morrisey2010preparing,varner2015mechanically}, and (\textit{ii}) side--branching, where ``daughter'' branches nucleate along the ``mother'' branch (far from the tip position)~\cite{yao2007matrix,varner2014cellular}. The interest here is in the mechanism of side--branching and, in particular, in the puzzling suppression of the side--branches, which has been observed through an excess of the matrix GLA protein (MGP) in the lungs vasculature~\cite{yao2007matrix}, as shown in Fig.~\ref{fig:jbc}. The experiments show that under excess MGP, which is the inhibitor of the bone morphogenetic protein (BMP), the nucleation of side--branches becomes sporadic as compared to the wild type. This phenomenon points toward a nonlinear nucleation mechanism, which may become essential upon the integration of biochemical and mechanic components~\cite{ochoa2012branching,hannezo2019multiscale,li2019multi}.
		\begin{figure}[tp]
			(a)\includegraphics[width=0.43\textwidth]{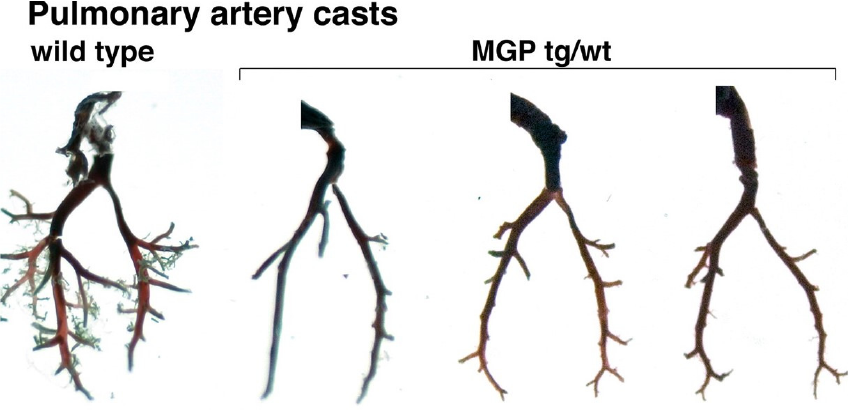}\vskip 0.1in
			(b)\quad\includegraphics[width=0.4\textwidth]{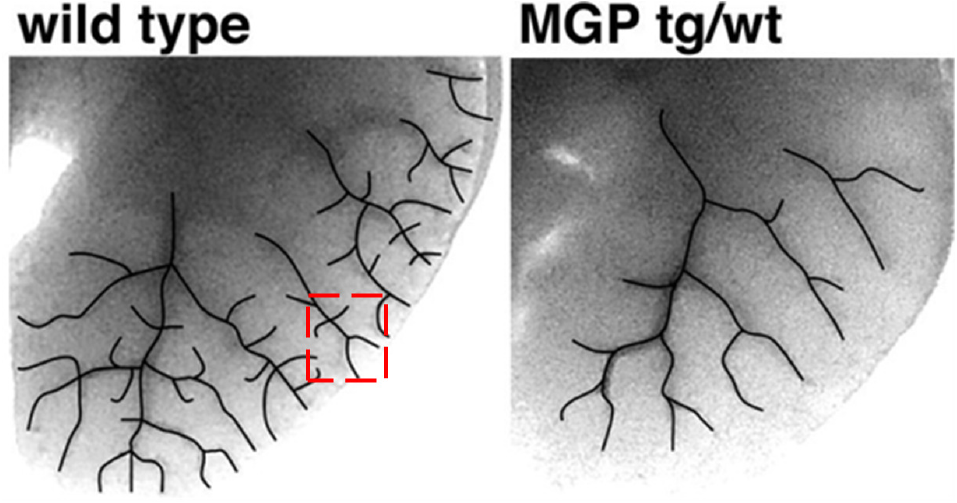}
			\caption{(a) Lung-vasculature morphology casts and (b) pulmonary vascular patterns from wild type littermates and MGP transgenic mice. The dashed-line rectangle in (b) demonstrates side--branching (dark lines), where segments evolve perpendicular to the main branch, in contrast to splitting that follows a Y-shaped dichotomy. This research was originally published in Y. Yao, S. Nowak, A. Yochelis, A. Garfinkel, and K. I. Bostr{\"o}m, Journal of Biological Chemistry, 282, 30131, 2007; licensed under a Creative Commons Attribution (CC BY) license.} 
			\label{fig:jbc}
		\end{figure}
		
		The mechanism of side--branching is a multiscale process~\cite{varner2017computational,hannezo2019multiscale,li2019multi} involving cues ranging from the molecular level to tissue compositions. A growing number of experiments indicate that biochemical signaling at the mesoscale dominates the initiation of side--branches~\cite{sainio1997glial,tang1998ret,lebeche1999fibroblast,tang2002ureteric,gilbert2004matrix,yao2007matrix,metzger2008branching,affolter2009tissue,yao2011matrix,hagiwara2015vitro,menshykau2019image}. In other words, similar {feedback loops, although in each developmental process these interactions are mediated by different proteins, were identified as operating in the development of the lungs, salivary gland, and kidney, see details in~\cite{affolter2009tissue,iber2013control}}. Traditionally, biochemical circuits point towards Turing's morphogenesis mechanism~\cite{turing1952chemical}. This mechanism underlies the interaction between molecules of activator and inhibitor substances and indeed has also been adopted to shed light on certain aspects related to the branching phenomenon, ranging from mammalian organs to plant roots~\cite{yao2007matrix,menshykau2019image,jilkine2007mathematical,payne2009theoretical,krupinski2016model,li2019multi,champneys2021bistability}. While numerical simulations show a similarity to the empirically observed dichotomy of branching mechanisms~\cite{metzger2008branching,hirashima2009mechanisms,menshykau2012branch,blanc2012role,celliere2012simulations,guo2014branching,guo2014mechanisms,xu2017turing,shan2018meshwork,zhu2018turing,menshykau2019image,guo2021wavelength}, the explicit instability mechanism of side--branching remains unclear~\cite{varner2017computational}. Therefore, it is of importance to clarify the role played by biochemical signaling in side--branching initiation and, especially, to address the pattern formation mechanism under excess inhibitor (e.g., MGP) as in~\cite{yao2007matrix}.
		
		Motivated by lung-vasculature development, I use an activator-inhibitor-substrate (AIS) model~\cite{yao2007matrix}, and reveal a distinct nonlinear nucleation mechanism. Analysis of the model suggests that spatially localized activator (BMP) peaks at the endothelial cells’ differentiation zone drive the initiation of new branches. The length scale depends on the domain size and on the nonlinear perturbations; near the Turing onset, however, the peaks may appear in ordered structure that resemble the classic Turing pattern. Due to the relatively cumbersome form of the model equations, the study mostly involves a bifurcation analysis via the numerical path continuation method in one spatial dimension (1D) and validations by direct numerical simulations (DNS) in 1D and 2D~\footnote{Direct numerical integrations have been performed using the commercial software COMSOL 5.2, with maximal element size of 0.1 and resolution of narrow regions of 0.02}. The mechanism belongs to the foliated homoclinic snaking universality class~\cite{knobloch2020stationary} and appears as robust. Furthermore, beyond significance to applications, the problem setting introduces new pattern selection mechanism that arise beyond the typically employed two-variable models, such as FitzHugh--Nagumo, Lugiato--Lefever, Gray--Scott, and Gierer--Meinhardt systems.
		
		\section{Model equations and bistability} 
		In 1976, Meinhardt ~\cite{meinhardt1976morphogenesis} proposed several AIS model equations to describe different aspects of the branching framework. One system has been employed to qualitatively tackle vascular and lung development~\cite{yao2007matrix}
		\begin{eqnarray}\label{eq:AI}
			\nonumber \frac{\partial A}{\partial t}&=&\frac{cSA^2}{H}-\mu A+\rA Y+D_{\text A} \nabla^2 A=F_A+D_{\text A} \nabla^2 A, \\
			\nonumber \frac{\partial H}{\partial t}&=&cSA^2-\nu H+\rH Y+D_{\text H} \nabla^2 H=F_H+D_{\text H} \nabla^2 H, \\
			\frac{\partial S}{\partial t}&=&c_0-\gamma S-\varepsilon Y S+D_{\text S} \nabla^2 S=F_S+D_{\text S} \nabla^2 S, \\ 
			\nonumber \frac{\partial Y}{\partial t}&=&d A-eY+\frac{Y^2}{1+fY^2}=F_Y,
		\end{eqnarray}
		where $A$, $H$, and $S$ are diffusible concentrations of the activator (BMP), inhibitor (MGP), and substrate (TGF-$\beta$/ALK1), respectively, while $Y$ represents an irreversible marker for differentiated endothelial cells. {The activator $A$ follows autocatalysis, also accompanied by a positive feedback from the substrate $S$, and inhibited by $H$. The activator triggers cell differentiation at rate $d$, representing a commitment to differentiate via the $Y$ field. For a detailed description of the biochemical signaling, the reader is referred to Yao \textit{et al.}, 2007~\cite{yao2007matrix}. For consistency, I also use here the rate of inhibitor secretion by cells, $\rH$, as a control parameter, while keeping all other parameters fixed, as in~\cite{yao2007matrix}: $c=0.002$, $\mu=0.16$, $\rA=0.005$, $\nu=0.04$, $c_0=0.02$, $\gamma=0.02$, $\varepsilon=0.1$, $d=0.008$, $e=0.1$, $f=10$, $D_{\text A}=0.001$, $D_{\text H}=0.02$, $D_{\text S}=0.01$.}
		
		\begin{figure}[b!]
			(a)\includegraphics[width=0.98\columnwidth]{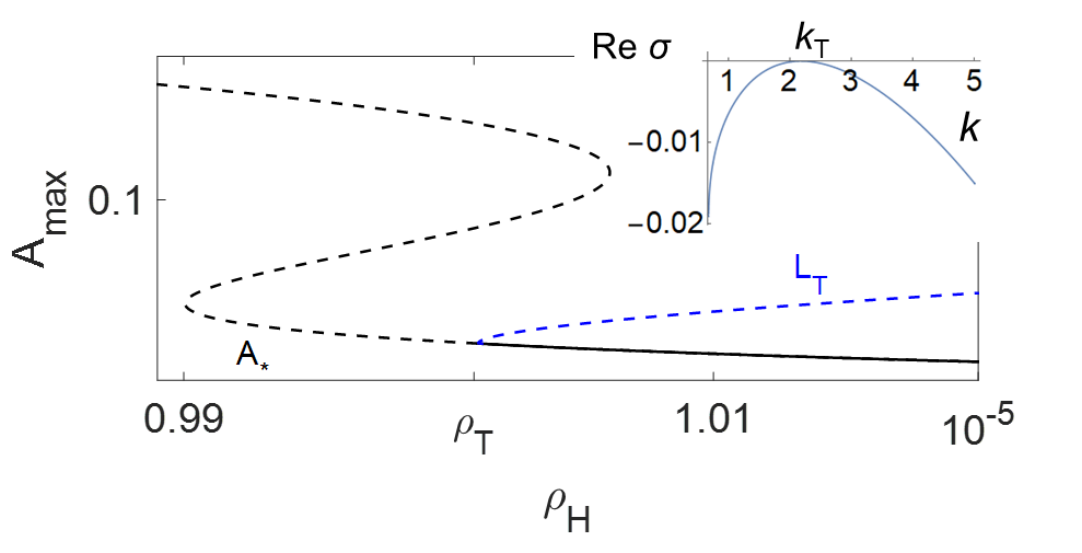}\\
			(b)\includegraphics[width=0.98\columnwidth]{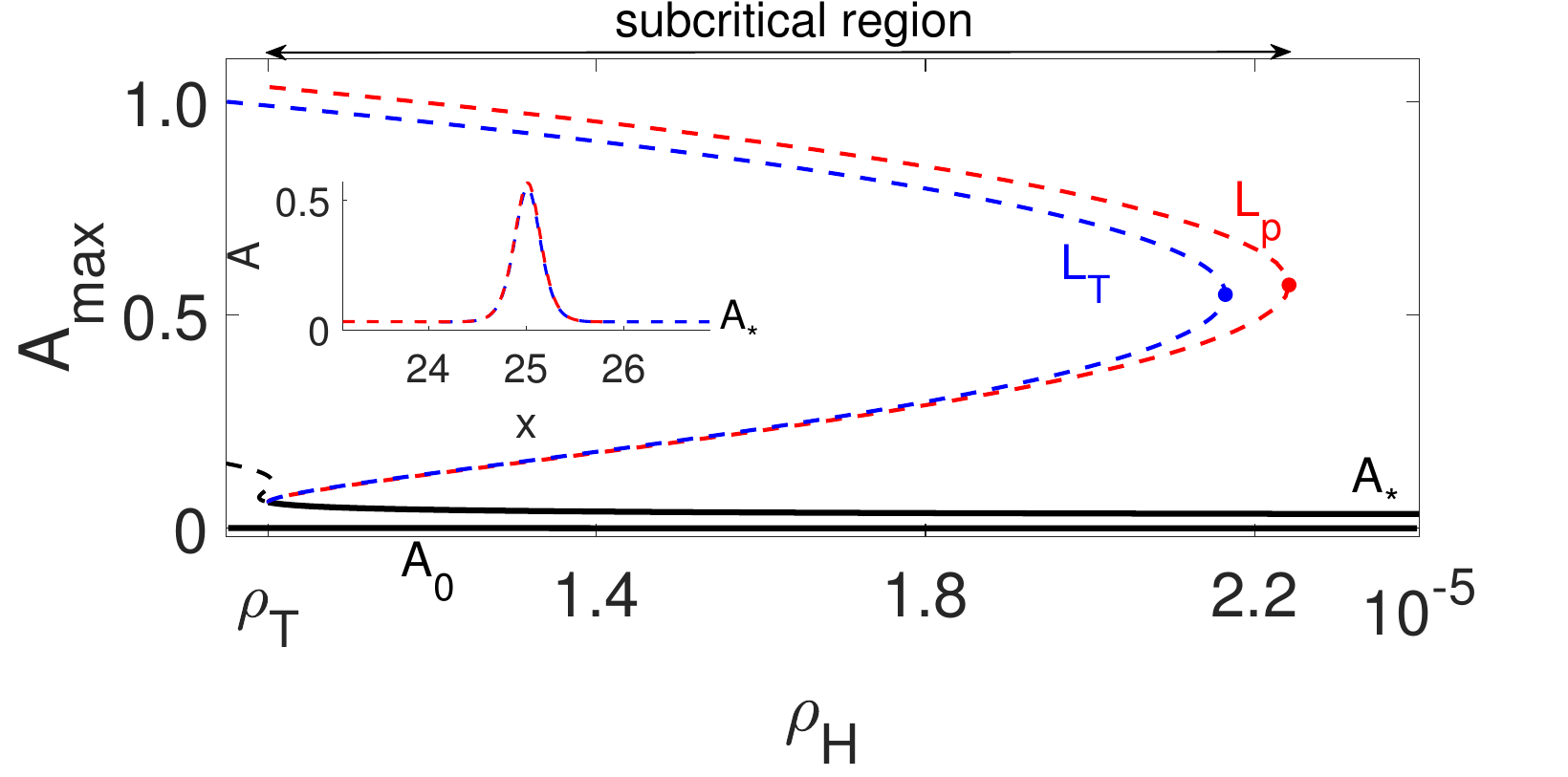}
			\caption{(a) Bifurcation diagram showing the activator values, $A$, the Turing instability onset in 1D, $\rT \simeq  1.0\cdot 10^{-5}$, and the bifurcating branch ($\LT$) of periodic (Turing) solutions (blue dashed line), where the inset depicts the dispersion relation at the onset following~\eqref{eq:turing}. The solid/dashed lines denote stable/unstable solutions. (b) Bifurcation diagram for uniform solutions $\vP_0$ and $\vP_*$, periodic (Turing) (on domain size of $\LT=2\pi/\kT\simeq 2.88$), and peak (on domain size $\LH=50$) solutions that bifurcate effectively from the Turing onset, $\rH = \rT$, computed via a path continuation~\cite{doedel2012auto} of~\eqref{eq:AIode}. The inset shows the respective profiles at the saddle--nodes that are marked by “$\bullet$,” respectively; for the peak solution (red line) only a portion of the $\LH$ domain is shown.} \label{fig:bif_uni}
		\end{figure}
		
		System~\eqref{eq:AI} has several uniform solutions: In addition to the “trivial” stable solution $\vP_0 \equiv (A_0,H_0,S_0,Y_0)=(0,0,c_0/\gamma,0)$, there are regions in which a multiplicity of nontrivial solutions coexist~\cite{knobloch2020stationary}. Of these, only one is linearly stable under uniform perturbations, and is referred to as $\vP_* \equiv (A_*,H_*,S_*,Y_*)$ (see Fig.~\ref{fig:bif_uni}); other solutions are not of interest here and thus are not shown. While $\vP_0$ is also linearly stable under nonuniform perturbations, $\vP_*$ goes through a Turing (finite wavenumber) instability~\cite{ch93}, where periodic perturbations with wavenumber $k$,
		\begin{equation}\label{eq:turing}
			\vP-\vP_* \propto e^{\sigma t + ikx},
		\end{equation}
		begin to grow exponentially at a rate $\sigma = \sigma(k) > 0$. The instability sets in at $\rH=\rT\simeq 1.0\cdot 10^{-5}$ and the critical wavenumber is $k=\kT~\simeq 2.18$; see inset in Fig.~\ref{fig:bif_uni}(a). 
		
		Examination of the experimental results~\cite{yao2007matrix}, however, does not indicate any undulations accompanying the differentiated ``mother'' branch or a typical length scale of the side--branches emanating from it (see Fig.~\ref{fig:jbc}). Thus, our interest is in the bistable region (coexistence of stable $\vP_*$ and $\vP_0$ states), $\rH>\rT$, in which a front solution exists, i.e., differentiated and non-differentiated endothelial cells. Additionally, previous numerical simulations show~\cite{yao2007matrix,guo2014branching,shan2018meshwork,zhu2018turing,varner2015mechanically,menshykau2019image,guo2021wavelength} another persistent feature that is related to the differentiation region, a strong local overshoot in expression of the activator that appears in all spatial dimensions, a characteristics that is consistent with experimental observations~\cite{weaver1999bmp,weaver2000bmp4,mailleux2001evidence,menshykau2019image}. 
		
		In what follows, it will be shown that this overshoot is, in fact, related to an isolated peak solution embedded in a
		background of uniform state $\vP_*$, whose role is twofold: To serve as a triggering event in a direction perpendicular to the tip motion and as an underlying pattern selection mechanism through which side--branches may appear. This phenomenon is summarized in DNS for a slightly perturbed (near the edges in $x$ direction) planar front-like initial condition, as shown in Fig.~\ref{fig:2Dsim}. For a high value of $\rH$ (corresponding to the excess MGP), the interface between $\vP_*$ and $\vP_0$ is stable with no peak formation (Fig.~\ref{fig:2Dsim}(a)), while for the lower $\rH$ value (wild type), yet above $\rT$, peaks do form not only at the perturbed locations but also in the middle of the domain and far from the initially created side--branches. Some of them then stabilize and form 2D propagating side--branches, as shown in Fig.~\ref{fig:2Dsim}(b). To understand this surprising pattern formation mechanism, I analyze~\eqref{eq:AI} in the context of nonuniform solutions and the role they play in applied perturbations.
		
		{\section{The emergence of peak solutions from the Turing onset}}
		For the existence of stationary peak solutions in~\eqref{eq:AI}, I exploit the spatial dynamics method by seeking solutions after rewriting~\eqref{eq:AI} as first-order differential equations, {where the time variable is replaced by the space variable}:
		\begin{eqnarray}\label{eq:AIode}
			A_x&=&-a, \, H_x=-h, \, S_x=-s, \, Y_x=-{z,} \\ 
			\nonumber a_x&=&F_A/D_{\text A},\, h_x=F_H/D_{\text H},\, s_x=F_S/D_{\text S},\,  {z}_x=F_Y/D_{\text Y}.
		\end{eqnarray}
		Note that an inconsequential weak diffusion for the $Y$ field ($D_{\text Y}=10^{-7}\ll D_{\text A}$) was added for numerical regularity.
		Stability is complemented by standard eigenvalue computations using a linearized version of~\eqref{eq:AI}. 
		\begin{figure}[tp]
			(a)\includegraphics[width=0.43\textwidth]{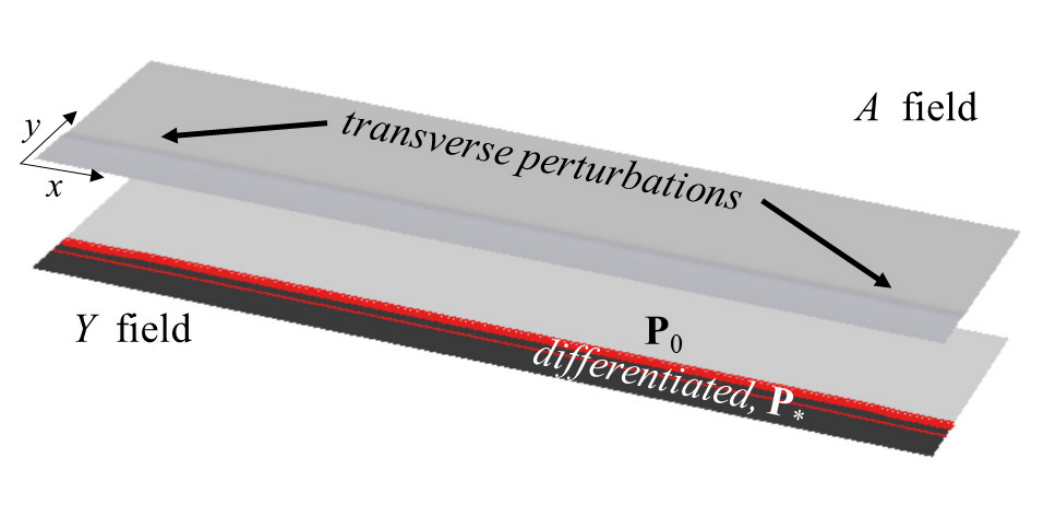}
			(b)\includegraphics[width=0.43\textwidth]{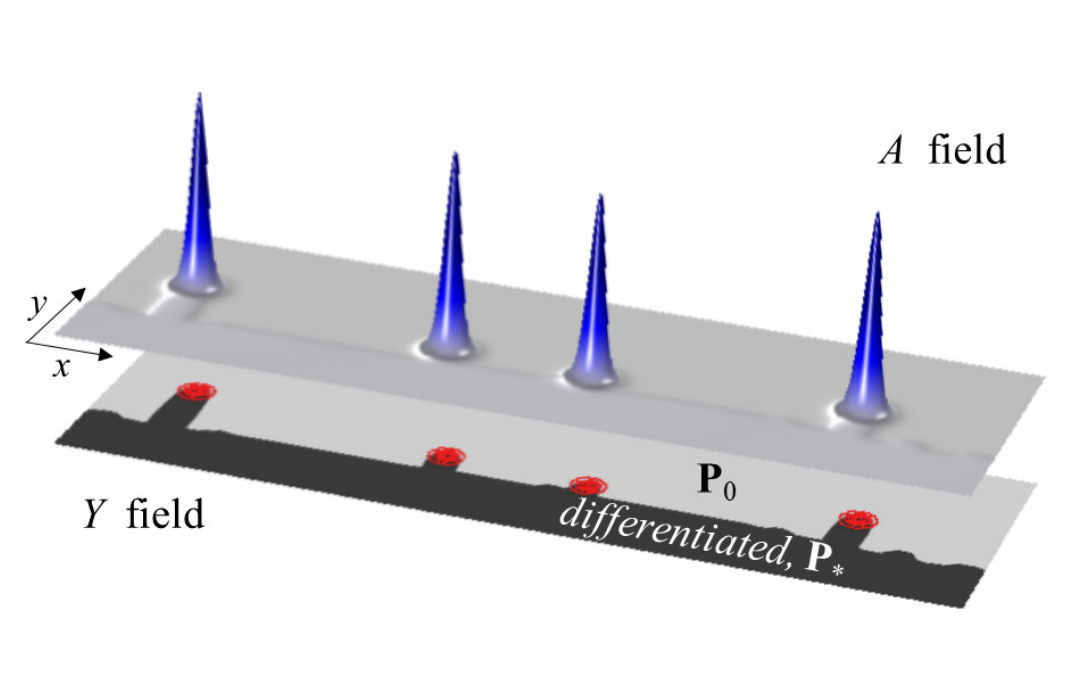}
			\caption{Snapshots of DNS of~\eqref{eq:AI} at $t=1000$ showing the activator field ($A$) on the top and the differentiation field ($Y$) on the bottom, as computed on a rectangular domain $[x,y]\in[16,5]$ with Neumann boundary conditions. The initial condition is an interface that connects $\vP_*$ and $\vP_0$, and dark colors indicate higher values of $A$ and differentiated $Y$ fields. The (red) contour lines in the bottom panel mark the locations of the localized $A$ states shown in the top panel. Parameters: (a) $\rH=3.0 \cdot 10^{-5}$ and (b) $\rH=2.0 \cdot 10^{-5}$. See supplementary movie (multimedia view) in 2D grey-scale for the time evolution of the $A$ field for (b). 
			} 
			\label{fig:2Dsim}
		\end{figure}
		
		Using~\eqref{eq:AIode}, it is possible to numerically obtain (using the path-continuation package AUTO~\cite{doedel2012auto}) the primary branch of bifurcating Turing solutions~\footnote{In computations using the AUTO package, spatial discretization typically followed: NTST=400 and NCOL=4}, i.e., solutions with periodicity $\LT=2\pi/\kT \simeq 2.88$, where $\kT$ is the critical wavenumber at the instability onset $\rH=\rT$. The periodic solutions, $\LT$, bifurcate subcritically, that is, towards the stable direction of $\vP_*$ (see Fig.~\ref{fig:bif_uni}), and are unstable. Typically, periodic solutions of the Turing type are accompanied by groups of nearby peaks~\cite{yochelis2008formation} that organize in the so-called ``snakes and ladders'' structure~\cite{burke2007snakes}. By performing a continuation on large domains, $\LH=50\gg \LT$, peaks are indeed found ($\LH$ is a periodic domain {whose} length can be as large as desired), but they are isolated and do not form the typical ``snakes and ladders'' structure. {Instead, these isolated peaks organize in a distinct \textit{foliated} homoclinic snaking structure, in which peaks repel each other. In region $\rT<\rH\lesssim 2.24\cdot 10^{-5}$, there are many additional coexisting solutions, but these are studied in more detail elsewhere~\cite{knobloch2020stationary}. Consequently, the peaks approach equidistant separation~\cite{yochelis2008formation,parra2018bifurcation}, which means that the length scale depends on the domain size and the nonlinear perturbations throughout the domain as opposed to the typical Turing patterns.} 
		
		The saddle-node (SN) bifurcation at which the peak solutions disappear (on the {branch} that is labeled as $\LH$) extends beyond the existence region of periodic Turing states ($\LT$), i.e., for larger values of $\rH$. The inset in Fig.~\ref{fig:bif_uni}(b) shows that the periodic and the peak solutions have the {similar} form, although $\LT \ll \LH$, as demonstrated via the profiles at both SN bifurcations; for the peak solution (red line) only part of the domain is shown. All periodic and peak solutions are linearly unstable to oscillations in 1D (where large amplitude solutions are additionally unstable to oscillations) while {in 2D peak solutions (which are in fact spots) become stable, even beyond the SN of the $\LH$ branch}; it is common that existence and stability regions differ from 1D to 2D~\cite{lloyd2008localized,gavish2017spatially}. This explains why, in 2D DNS, the peaks are locked to the differentiation-front region. Temporal stability will be studied in detail, elsewhere.
		\begin{figure}[tp]
			(a)\includegraphics[width=0.2\textwidth]{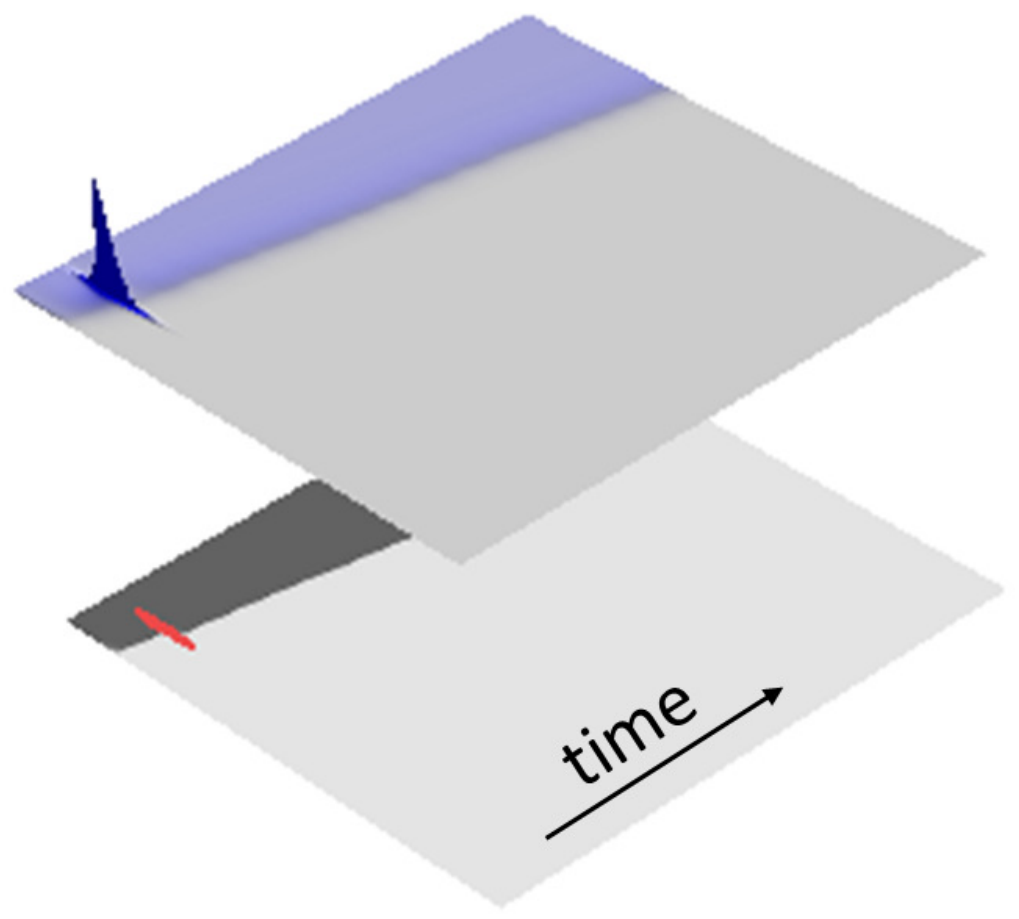}
			(b)\includegraphics[width=0.2\textwidth]{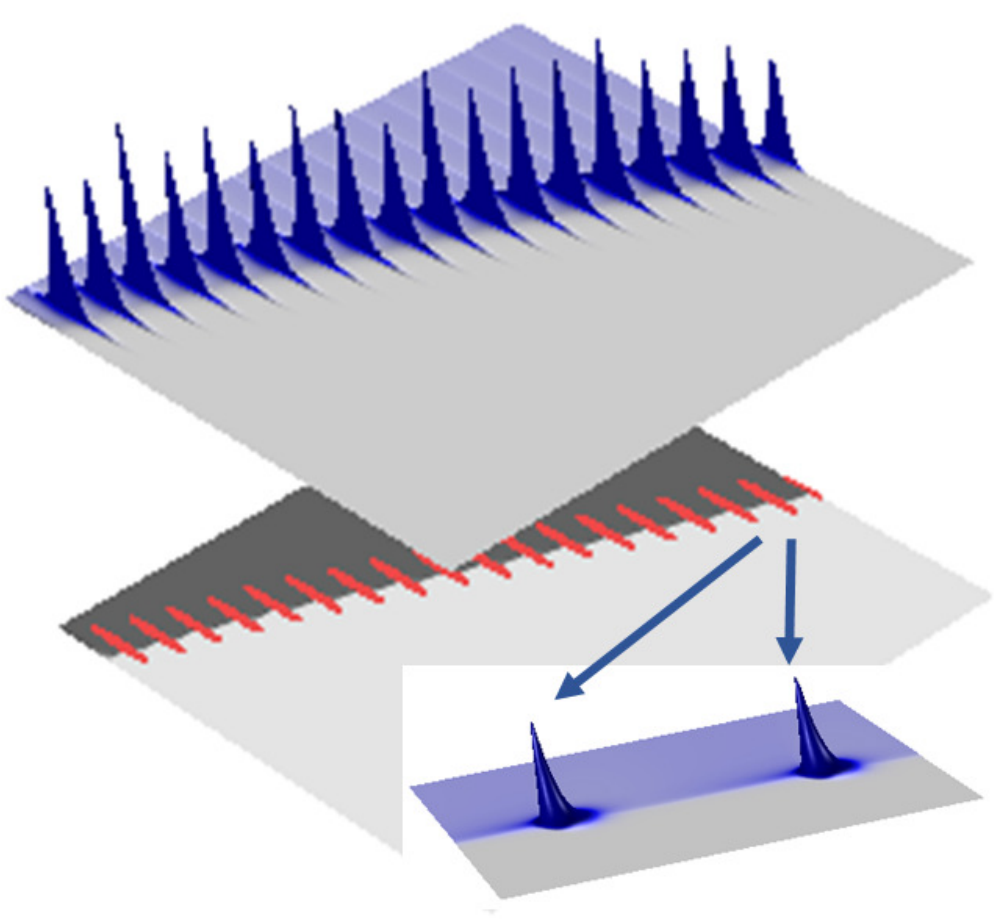}\vskip 0.1in
			(c)\includegraphics[width=0.42\textwidth]{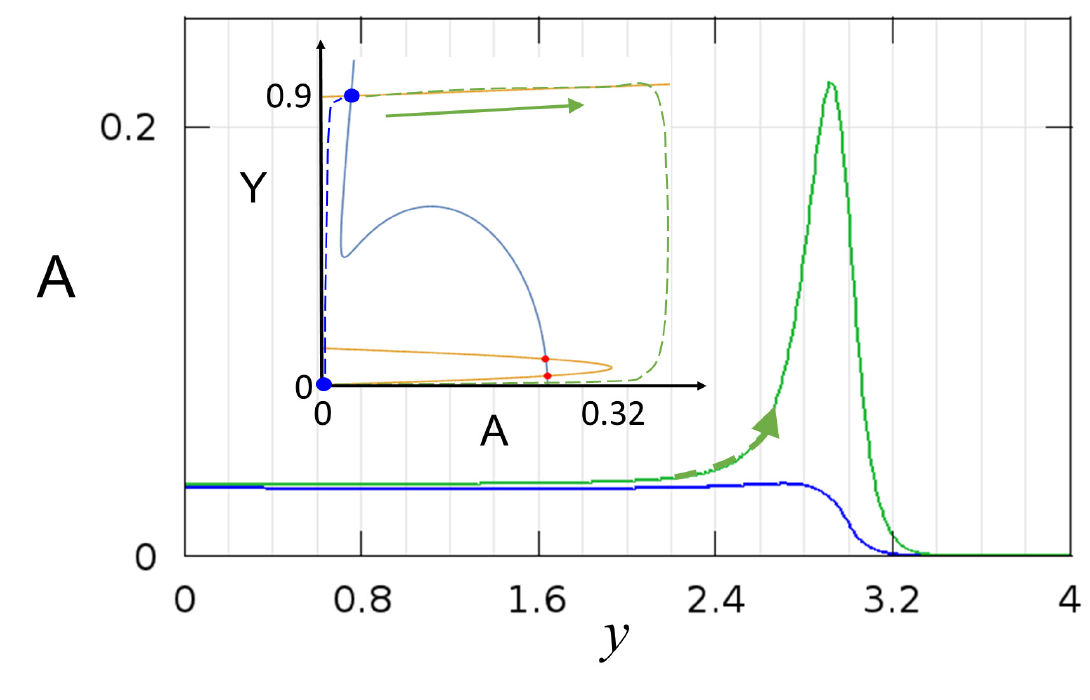}
			\caption{Space-time plots of~\eqref{eq:AI} showing both the activator $A$ (top panel) and cell differentiation $Y$ (bottom panel) fields for: (a) $\rH=3.0\cdot 10^{-5}$ and (b) $\rH=2.0\cdot 10^{-5}$, where nonuniform solutions coexist. The contour lines in the bottom panel mark the locations of the localized $A$ states shown in the top panel. DNS were performed on a spatial domain $y \in [0, 5]$ and times $t \in [0, 3000]$; dark color indicates higher values of both fields. Boundary conditions are Neumann, and the initial condition is a front connecting $\vP_*$ and $\vP_0$ at $y=0.5$. The inset in the bottom panel shows a close-up of about a single oscillation period for $t\in [2600, 2900]$, as also indicated by the arrows. (c) Two front profiles are taken from Fig.~\ref{fig:SW}(b), showing a solution before peak formation (blue line, time 2910) and at the beginning of peak formation (green line, time 2996). The inset shows the nullclines of~\eqref{eq:AI} in the $A-Y$ plane, marking the unstable fixed points by red “\colr{$\bullet$}” and stable fixed points by blue “\colb{$\bullet$}.” The heuristic front trajectory is depicted by the dashed lines, where the left trajectory corresponds to the bottom profile (blue profile); the arrow indicates the direction of the peak as also indicated in the main figure by the dashed arrow in the top profile (green color).} 
			\label{fig:SW}
		\end{figure}
		
		\section{Nonlinear transverse front instability and side--branching}
		
		The existence of peak solutions in 1D triggers the side--branching nucleation in 2D (see Fig.~\ref{fig:2Dsim} and the supplementary movie (multimedia view)) by two processes that act simultaneously in the differentiated front $y$ direction and in the transverse to the front line direction, i.e., in the longitudinal $x$ direction. These two processes underline a distinct nonlinear transverse front instability for bistable systems. For convenience, we discuss the impacts of these two processes separately by referring to the former as the case (\textit{i}) and to the latter as the case (\textit{ii}).
		
		{Case (\textit{i}) is related to the differentiation process that exhibits qualitatively different behaviors if the parameters are chosen to be at or outside the subcritical region, as summarized in Fig.~\ref{fig:SW}. For high values of $\rH$ that are outside the peak existence region, the differentiation front is linearly stable as demonstrated by a space-time plot, showing that after a transient overshoot at early times, the front propagates without any structural changes (Fig.~\ref{fig:SW}(a)). Stability of the front in 1D (in $y$ direction) also results in stability of planar fronts in 2D and therefore, side--branches do not form, which is consistent with the DNS demonstrated in Fig.~\ref{fig:2Dsim}(a). For $\rH$ values that are inside the subcritical Turing regime (Fig.~\ref{fig:bif_uni}(b)), fronts are unstable and their propagation involves strong peak oscillations at the front line, as shown in Fig.~\ref{fig:SW}(b). The fluctuations in the peak amplitude are attributed to relatively sharp oscillations (see inset in bottom panel in Fig.~\ref{fig:SW}(b)) that are difficult to resolve on discrete grids in space and time before reaching asymptotic oscillations (not shown here). The locked peak oscillations at the propagating front line can be understood through the AIS kinetics using the dynamics along the nullclines, i.e., in the $(A,Y)$ phase plane with invariant manifolds $F_A=0$ and $F_Y=0$. Figure~\ref{fig:SW}(c) depicts two typical front profiles in the $(A,Y)$ phase plane, the first at times before peak formation (bottom blue line) while the second through the peak formation (top green line), the inset shows the nullclines (solid blue/yellow lines) together with the heuristic trajectories (dashed blue/green lines, respectively) of the profiles. 
			
			The front is a nonlinear perturbation as it connects two uniform states (i.e., it is a heteroclinic connection) so that any local fluctuation in the front line, such as propagation, increases the $A$ field due to an abrupt increase in the $Y$ field. Therefore, if the $\rH$ value is within the subcritical region, the trajectory follows the top nullcline (as indicated by the arrow in the inset) along the manifold of the peak solution, which is unstable. Since the peak solutions are of order one in their amplitudes, their height is above $A\simeq 0.5$ (see inset in Fig.~\ref{fig:bif_uni}(b)), the trajectory makes a large excursion (the overshoot) before connecting to the trivial state $\vP_0$ at $(A,Y)=(0,0)$. Consequently,  the ``flow'' about the unstable peak manifold (in $y$ direction) provides a robust symmetry breaking mechanism through which nonuniform perturbations along the transverse front line (in $x$ direction) lead to the formation of stable localized spots (side--branching stems in Fig.~\ref{fig:2Dsim}(b)) out of the planar rim oscillations, as shown in the supplementary movie (multimedia view). Notably, this trajectory is absent to the right of the subcritical regime and thus, in the absence of peak solutions, side--branching is suppressed, as shown in Fig.~\ref{fig:2Dsim}(a).}
		\begin{figure}
			(a)\includegraphics[width=0.13\textwidth]{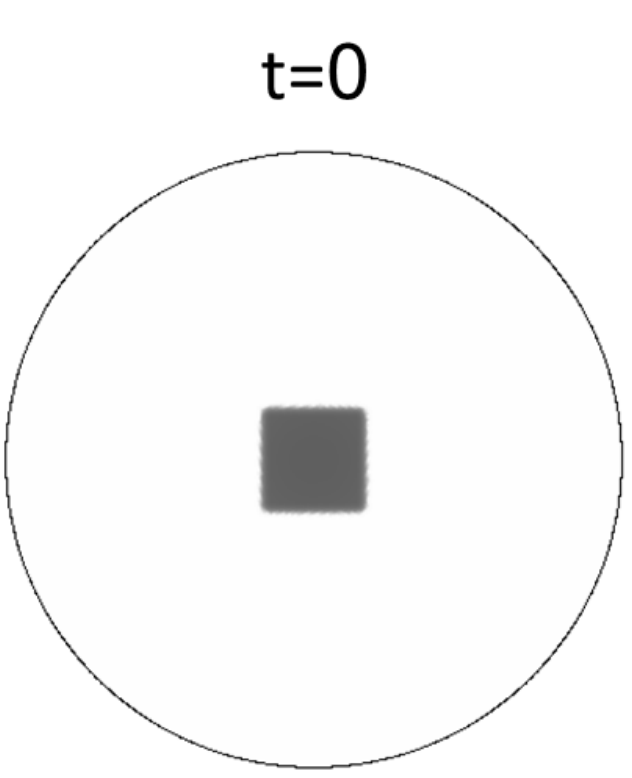}
			(b)\includegraphics[width=0.13\textwidth]{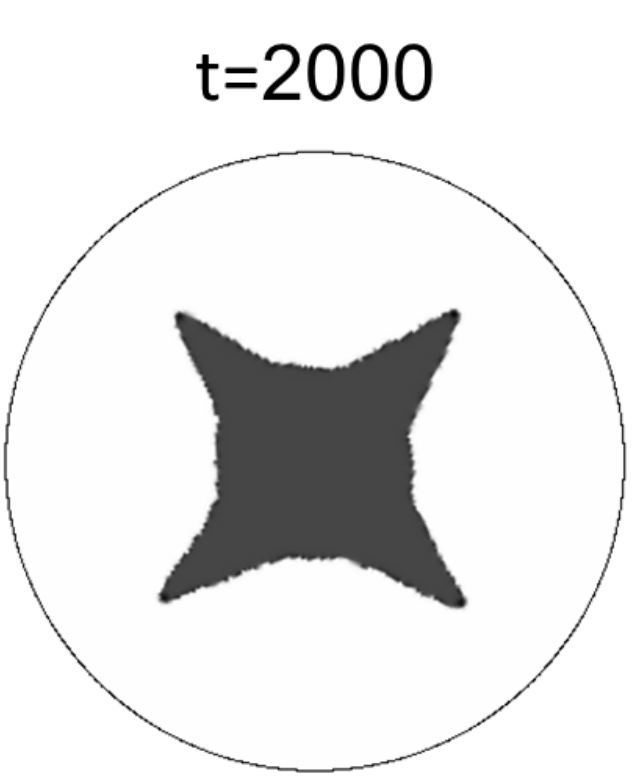}
			(c)\includegraphics[width=0.13\textwidth]{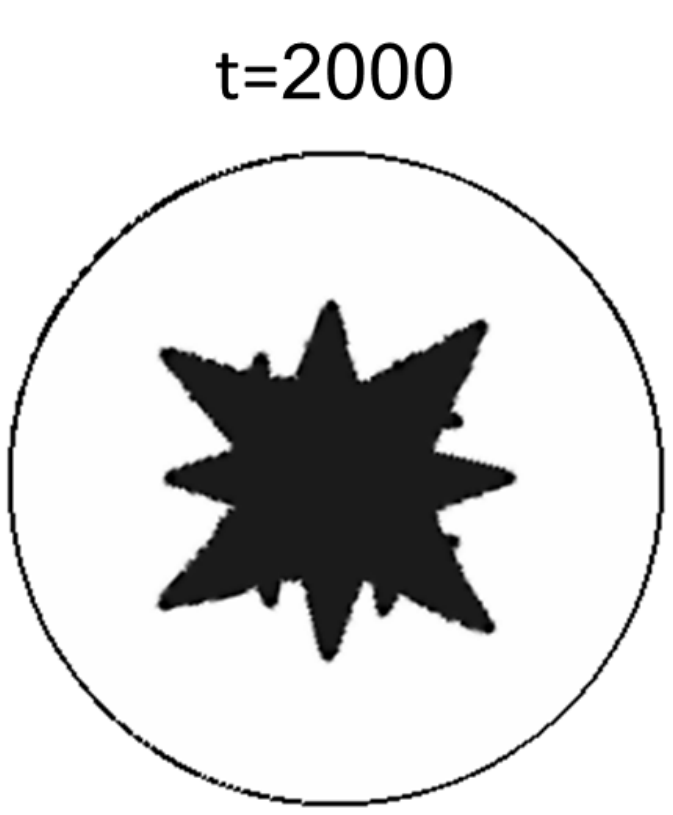}
			\caption{Snapshots of DNS of~\eqref{eq:AI} showing the cell differentiation field, $Y$, computed on a circular domain of diameter 30 and with Neumann boundary conditions, where the initial condition is depicted in (a); dark color indicates higher values of $Y$. {Notably, images (b) and (c) are not asymptotic solutions.} Parameters: (b) $\rH=3.0\cdot 10^{-5}$, (c) $\rH=2.0\cdot 10^{-5}$.} \label{fig:branching}
		\end{figure}
		
		{Case (\textit{ii}) is related to the wavenumber selection of the nucleated peaks along the transverse front line (in $x$ direction) and is associated with the repulsive nature of the formed side--branches that are driven by the nature of the spatially localized perturbations. This stems from the foliated homoclinic organization of peaks in 1D~\cite{knobloch2020stationary} and thus, results in the formation of isolated side--branches, in which separation distances depend on the applied perturbations. In DNS shown in Fig.~\ref{fig:2Dsim}, I used weak (yet finite) transverse perturbations along the front line that indeed in the subcritical regime, resulted in formation of side--branches. However, the stability of spot solutions in 2D (as opposed to 1D) implies that large localized perturbations may also trigger side--branches. In biological context, such perturbations can be attributed to mechanical instabilities~\cite{blanc2012role,varner2015mechanically,kourouklis2018modeling} or relatively localized biochemical stimuli, such as Fgf10~\cite{hirashima2009mechanisms,affolter2009tissue} and the VEGF family~\cite{mettouchi2012role}.}
		
		{To test the roles of large perturbations, I performed DNS using circular domains and with a square shape initial condition that has well-defined high curvature regions (Figs.~\ref{fig:branching}(a)), i.e., strong localized (nonlinear) perturbations. Figure~\ref{fig:branching}(b) shows that for high $\rH$ values that are beyond the subcritical regime, the side--branches form only from the corners of the initial condition. This behavior implies that to the right of the subcritical region only sporadic side--branches can form, a situation that resembles excess MGP that is shown in Fig.~\ref{fig:jbc}(b), i.e., the side--branches are not expected to have any typical length scale. On the other hand, if $\rH$ is in the subcritical region, side--branches form also along the planar portions, in accordance with case (\textit{i}). This equidistant space filling patterning implies that in this regime, side--branches are likely to form a defined periodicity (see Fig.~\ref{fig:jbc}(a)) that may be confused with a classical Turing mechanism~\cite{guo2014mechanisms,xu2017turing,zhu2018turing,menshykau2019image,guo2021wavelength}.} 
		
		Notably, the subcritical nature of nonuniform solutions explains the (artificial) need for parameter fluctuations as nonlinear perturbations that have been employed in previous DNS to trigger side--branching~\cite{meinhardt1976morphogenesis}, i.e., fluctuations in the value of $c_0$ at each time step~\cite{yao2007matrix,guo2014mechanisms}. Here it is shown that such fluctuations are not essential.
		
		\section{Discussion}
		{By using a bifurcation analysis of an activator-inhibitor-substrate system in the bistable regime of homogeneous states, it is possible to reveal a distinct nonlinear transverse front mechanism that may explain the different forms of side--branching development, such as in the case of pulmonary vascular pattern~\cite{yao2007matrix}, as shown in Fig.~\ref{fig:jbc}. Owing to repelling peaks that emerge subcritically from the Turing onset, the nucleation mechanism of side--branches involves a transverse front instability that is different from other linear and nonlinear fingering instabilities in bistable reaction--diffusion type systems~\cite{hagberg1994labyrinthine,goldstein1996interface,yochelis2004two,hagberg2006linear}, not only by the presence of the oscillatory phase that precedes the break up into fingers but also in the absence of a typical length scale. The pattern formation mechanism appears as robust since it is related to global bifurcations, i.e., homoclinic and heteroclinic connections in space describing peaks and fronts, respectively. In the context of pulmonary vascular pattern, the results suggest that the excess amount of the inhibitor (MGP) corresponding to high values of $\rH$, suppresses in general the formation of side--branches. However, in the presence of large nonlinear perturbations, such as mechanical instabilities or spatially localized biochemical stimuli~\cite{hirashima2009mechanisms,affolter2009tissue,varner2015mechanically,kourouklis2018modeling,menshykau2019image,li2019multi}, side-- branches may still sporadically form, as shown in Fig.~\ref{fig:jbc}(b).} 
		
		More broadly, similar nonlinear transverse instability is expected to be a generic feature of other systems, where isolated peaks along with bistability have been reported, such as plant root-hairs~\cite{brena2014mathematical,draelants2015localized}, transition of epithelial/mesenchymal phenotypes to metastasis~\cite{glienke2000differential,lee2012epithelial,jolly2015implications,garg2017epithelial,liao2020hybrid}, formation of filopodia by actin polymerization~\cite{isaac2013linking,szymanski2018actin,fischer2019filopodia}, and nonlinear optics~\cite{parra2018bifurcation}. 
		
		\begin{acknowledgments}
			I thank Edgar Knobloch (UC Berkeley) and Ehud Meron (BGU) for insightful discussions, and I am also grateful to the anonymous referees whose comments significantly improved the clarity of the paper.
		\end{acknowledgments}
		
		\section*{DATA AVAILABILITY}
		The data that support the findings of this study are available from the corresponding author upon reasonable request.
		
		\section*{REFERENCES}

	\end{document}